\documentclass[12pt]{article}
\usepackage{authblk}
\usepackage[bookmarksnumbered, colorlinks, plainpages]{hyperref}
\usepackage{amsmath, amsthm, amscd, amsfonts, amssymb, graphicx, color, booktabs,enumitem}
\numberwithin{equation}{section}
\definecolor{email}{rgb}{0.00,0.00,0.84}
\begin{document}
\setcounter{page}{1}

\title{\large \bf 12th Workshop on the CKM Unitarity Triangle\\ Santiago de Compostela, 18-22 September 2023 \\ \vspace{0.3cm}
\LARGE $C\!P$ violation in $D$ meson decays at Belle I/II}

\author[1]{Michel Bertemes on behalf of the Belle I/II collaboration}
\affil[1]{Brookhaven National Laboratory (BNL), Upton,
NY 11973-5000, USA}
\maketitle

\begin{abstract}
We present recent measurements from the Belle and Belle II experiment related to $CP$ violation in decays of charmed mesons via two complementary approaches. We also propose a new algorithm to determine the flavor of neutral charmed mesons.
\end{abstract} \maketitle

\section{SuperKEKB and Belle II}

SuperKEKB is an asymmetric $e^+e^-$-collider located in Tsukuba, Japan. The respective beam energies are 7 GeV ($e^-$) and 4 GeV ($e^+$), resulting in a centre-of-mass energy tunable around the $\Upsilon(4S)$ mass. Thanks to the nano-beam collision scheme, a record instantaneous luminosity of $4.7\times10^{34}\mathrm{cm}^{-2}\mathrm{s}^{-1}$ was achieved. Located at the interaction point of both beams is the Belle II experiment, a $4\pi$ spectrometer. Belle II is the successor to the Belle experiment with improved vertexing, tracking, calorimetry and particle identification capabilities. Belle II has been operated since 2019 and collected a dataset corresponding to an integrated luminosity of 424$\mathrm{fb}^{-1}$ (about half of what Belle collected from 1999-2010) at or above the $\Upsilon(4S)$ mass. After a shutdown in 2022 for detector and accelerator improvements, data taking resumed in 2024. Due to the large $e^+e^- \to c\bar{c}$ production cross-section, Belle II is recording approximately 1.3M charm events per 1$\mathrm{fb}^{-1}$ with a trigger efficiency uniform across decay time and kinematics. With an excellent reconstruction of final states with neutral particles, the charm physics program evolves around measurements of charge-conjugation and parity ($C\!P$) asymmetries and mixing of decays including $D^+ \to \pi^+\pi^0$ and $D^0 \to \pi^0\pi^0,\rho^0\gamma,K_S^0K_S^0,K\pi\pi^0,\pi\pi\pi^0$.

\section{Charm Flavor Tagger}

The measurements outlined above usually require the identification of the flavor of the neutral charmed hadron at the time of production. One way to accomplish the flavor tagging is by selecting neutral $D^0$ mesons that emerge from the strong decay $D^{*+}\to D^0 \pi^+$, where the charge of the pion determines the flavor of the $D^0$ meson. As the resulting number of $D^0$ mesons is much smaller than the total amount of $D^0$ mesons produced in $e^+e^-$ collisions, only a limited sample of $D^0$ mesons is available for measurements. 

The Charm Flavor Tagger (CFT) is a new algorithm to determine the production flavor of $D^0$ mesons \cite{cft}. It exploits the correlation between the flavor of a reconstructed $D^0$ meson and the electric charges, flavor and kinematic properties of particles reconstructed in the rest of the event. Included among the latter are particles from the decay of the other charmed hadron in the event and possibly also particles produced alongside the $D^0$ meson. The CFT only considers the tracks most collinear with the reconstructed meson, as these are most likely to tag the correct flavor. A binary classification algorithm is used to make a prediction of $qr$, the product of the tagging decision $q$ (+1 for $D^0$ mesons and -1 for $\bar{D^0}$ mesons) and the dilution $r$ ($r=1$ corresponding to a perfect prediction, $r=0$ being equivalent to random guessing). Different quantities related to kinematics and particle identification probabilities are used as input to the algorithm.

\begin{figure} [hbt!]
\centering
\includegraphics[width=0.55\textwidth]{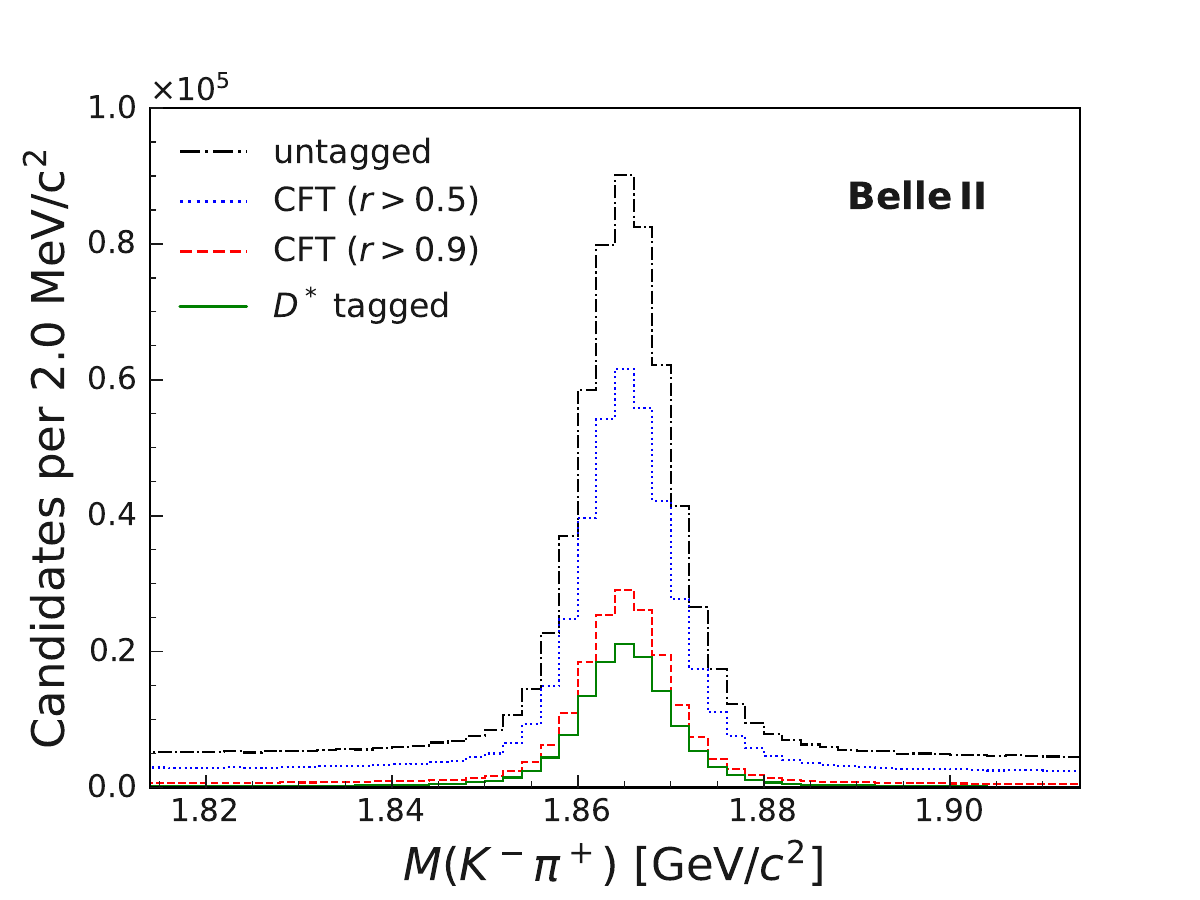}
 \caption{Mass distribution for $D^0\to K^- \pi^+$ decays with different requirements on the predicted dilution in comparison with $D^{*+}$-tagged decays. \cite{cft}}
\label{fig:cft}
\end{figure}

The CFT is trained using simulation and calibrated with data collected by the Belle II experiment corresponding to an integrated luminosity of 362$\mathrm{fb}^{-1}$. The tagging power, defined as $\epsilon_{\mathrm{tag}}^{\mathrm{eff}}=\epsilon_{\mathrm{eff}}\langle r^2\rangle$ ($\epsilon_{\mathrm{eff}}$ being the tagging efficiency), represents the effective sample size when a tagging decision is needed. We obtain:
\begin{equation}
    \epsilon_{\mathrm{tag}}^{\mathrm{eff}} = [47.91\pm0.07(\mathrm{stat})\pm0.51(\mathrm{syst})]\%
\end{equation}
This value is found to be independent of the reconstructed $D^0$ decay mode. The CFT is expected to double the effective sample size with respect to measurements that so far have relied exclusively on $D^{*+}$-tagged events. This increase also affects the amount of contributing background events (see Figure \ref{fig:cft}).

\section{$C\!P$ violation measurements}

Direct $C\!P$ violation can be searched for in measurements of time-integrated asymmetries from differences in partial widths:

\begin{equation}
    A_{C\!P}=\frac{\Gamma(D\to f) - \Gamma(\bar{D}\to \bar{f})}{\Gamma(D\to f) + \Gamma(\bar{D}\to \bar{f})}
\end{equation}
where $f$ and $\Bar{f}$ are $C\!P$-conjugate final states. The related experimental observable however also includes asymmetries in production and reconstruction which need to be subtracted with means of a control channel.

A complementary approach is based on the measurement of triple products $C_T = \Vec{\nu_1}\cdot(\Vec{\nu_2}\times\Vec{\nu_3})$, where $\Vec{\nu_i}$ represents the momentum of the final state particles. $C_T$ changes sign under time-reversal $T$ and if it is found to be asymmetric around zero, this could indicate $T$-violation. Under the $C\!P\!T$ symmetry, this could then constitute a $C\!P$ violation signal. The corresponding asymmetries for $D$ and $\Bar{D}$ are defined as:

\begin{equation}
    A_T = \frac{\Gamma(C_T>0)-\Gamma(C_T<0)}{\Gamma(C_T>0)+\Gamma(C_T<0)} ,\quad \Bar{A_T} = \frac{\Gamma(-\Bar{C_T}>0)-\Gamma(-\Bar{C_T}<0)}{\Gamma(-\Bar{C_T}>0)+\Gamma(-\Bar{C_T}<0)}
\end{equation}
where the minus sign corresponds to parity transformation which is required for $\Bar{A_T}$ to be the CP conjugate of $A_T$. 
Since $A_T\neq 0$ can also arise from final-state interactions, $T$-violation can be isolated by defining:
\begin{equation}
    a_{CP}^{T-\mathrm{odd}} = \frac{1}{2}(A_T-\Bar{A_T})
\end{equation}
$a_{CP}^{T-\mathrm{odd}}$ is unaffected by production and reconstruction asymmetries. In charm, four-body decays are well suited to study these asymmetries as they have sizable branching fractions and give access to three non-trivial momentum values.

These two approaches differ in their dependence on the strong phase $\delta$: While the direct asymmetries require $\delta\neq 0$ since $A_{CP}\propto \sin(\phi)\sin(\delta)$, this does not hold for measurements with triple products as $a_{CP}^{T-\mathrm{odd}}\propto \sin(\phi)\cos(\delta)$.

Three recent measurements are presented in the following, all based on the data set collected by the Belle experiment on or near the $\Upsilon(nS)$ ($n=1,2,3,4,5$) resonances, corresponding to an integrated luminosity of $\sim 1 \mathrm{ab}^{-1}$. $K_S^0$ candidates are identified with a neural network using kinematic features as input \cite{ksks,neuro}. Further background suppression is achieved by fitting the individual decay chains and applying a selection on the quality of the vertex fit as well as the flight-length significance of the $D$ meson.

\begin{enumerate}
    \item \textit{Measurement of the branching fraction and search for $C\!P$ violation in $D^0\to K^0_S K^0_S \pi^+\pi^-$ decays at Belle} \cite{aman}
\end{enumerate}
$C\!P$ violation is searched for in the Cabibbo-suppressed (CS) decay $D^0\to K^0_S K^0_S \pi^+\pi^-$ with the two approaches outlined above and by determining the flavor of the $D^0$ meson with a $D^*$-tag. First, the branching fraction is measured relative to that of  $D^0\to K^0_S \pi^+\pi^-$:
\begin{equation*}
    \mathcal{B}(D^0\to K^0_S K^0_S \pi^+\pi^-) = [4.79\pm0.08(\mathrm{stat})\pm0.10(\mathrm{syst})\pm0.31(\mathrm{norm})]\times 10^{-4} 
\end{equation*}
where the last uncertainty is due to $\mathcal{B}(D^0\to K^0_S \pi^+\pi^-)$. 
The time-integrated CP asymmetry is measured to be:
\begin{equation*}
    A_{CP}(D^0\to K^0_S K^0_S \pi^+\pi^-) = [-2.51\pm1.44(\mathrm{stat})^{+0.11}_{-0.10}(\mathrm{syst})]\% 
\end{equation*}
where the asymmetry related to the reconstruction of the soft pion $\pi_s^+$ has been determined from a study of flavor-tagged $D^{*+}\to D^0(\to K^-\pi^+) \pi_s^+$ and untagged $D^0 \to K^-\pi^+$ decays.
\begin{equation*}
    a_{CP}^{T-\mathrm{odd}}(D^0\to K^0_S K^0_S \pi^+\pi^-) = [-1.95\pm1.42(\mathrm{stat})^{+0.14}_{-0.12}(\mathrm{syst})]\% 
\end{equation*}
The branching fraction measurement is the most precise to date. The $C\!P$ measurements are the first ones for this decay. The largest systematic effect for all three measurements arises from background where the $D^*$ candidate is correctly reconstructed but a particle was missed in the reconstruction of the $D^0$ candidate. The amount is estimated with simulation. 

\begin{enumerate}[resume]
    \item \textit{Search for $C\!P$ violation in $D_{(s)}^+\to K^+ K^0_S h^+h^-$ ($h=K,\pi$) decays and observation of the Cabibbo-suppressed decay $D_{s}^+\to K^+ K^-K^0_S \pi^+$} \cite{moon}
\end{enumerate}
We report the first observation of the CS suppressed decay $D_s^+\to K^+ K^- K^0_S \pi^+$ and measure its branching fraction to be:
\begin{equation*}
    \mathcal{B}(D_s^+\to K^+ K^- K^0_S \pi^+) = [1.29\pm0.14(\mathrm{stat})\pm0.04(\mathrm{syst})\pm0.11(\mathrm{norm})]\times 10^{-4} 
\end{equation*}
where the last uncertainty is due to the uncertainty of the normalization channel $\mathcal{B}(D_s^+\to K^+ K^0_S \pi^+ \pi^-)$ and the largest systematic effect arises from the modeling of the signal shape. We measure $a_{CP}^{T-\mathrm{odd}}$ in the remaining channels:
\begin{align*}
    a_{CP}^{T-\mathrm{odd}}(D^+\to K^+ K^0_S \pi^+\pi^-) &=& [0.34\pm0.87(\mathrm{stat})\pm0.32(\mathrm{syst})]\% \\
    a_{CP}^{T-\mathrm{odd}}(D_s^+\to K^+ K^0_S \pi^+\pi^-) &=& [-0.46\pm0.63(\mathrm{stat})\pm0.38(\mathrm{syst})]\% \\
    a_{CP}^{T-\mathrm{odd}}(D^+\to K^+ K^- K^0_S \pi^+) &=& [-3.34\pm2.66(\mathrm{stat})\pm0.35(\mathrm{syst})]\% 
\end{align*}
The results are the most precise to date and consistent with no $C\!P$ violation. The biggest systematic uncertainty is related to a possible detector bias that we study with the Cabibbo-favored decay $D^+ \to K_S^0 \pi^+ \pi^+ \pi^-$ where $a_{CP}^{T-\mathrm{odd}}$ is expected to be consistent with 0.
\begin{enumerate}[resume]
    \item \textit{Search for $C\!P$ violation using $T$-odd correlations in $D_{(s)}^+\to K^+ K^- \pi^+ \pi^0$ $D_{(s)}^+ \to K^+ \pi^- \pi^+ \pi^0$ and $D^+\to K^- \pi^+ \pi^+ \pi^0$ decays} \cite{llki}
\end{enumerate}
A search for $C\!P$ violation is performed in five $D^+_{(s)}$ decays:
\begin{align*}
    a_{CP}^{T-\mathrm{odd}}(D^+\to K^- K^+ \pi^+\pi^0) &=& [0.26\pm0.66(\mathrm{stat})\pm0.13(\mathrm{syst})]\% \\
    a_{CP}^{T-\mathrm{odd}}(D^+\to K^+ \pi^- \pi^+\pi^0) &=& [-1.30\pm4.20(\mathrm{stat})\pm0.10(\mathrm{syst})]\% \\
    a_{CP}^{T-\mathrm{odd}}(D^+\to K^- \pi^+ \pi^+ \pi^0) &=& [0.02\pm0.15(\mathrm{stat})\pm0.08(\mathrm{syst})]\% \\
    a_{CP}^{T-\mathrm{odd}}(D_s^+\to K^+ \pi^- \pi^+ \pi^0) &=& [-1.10\pm2.20(\mathrm{stat})\pm0.10(\mathrm{syst})]\% \\
    a_{CP}^{T-\mathrm{odd}}(D_s^+\to K^- K^+ \pi^+ \pi^0) &=& [-0.22\pm0.33(\mathrm{stat})\pm0.43(\mathrm{syst})]\%
\end{align*}
where the largest systematic effect is due to a possible $C_T$-dependence of the reconstruction efficiency, which is studied with signal simulation samples. The results are all consistent with zero and show no evidence of $C\!P$ violation. These are the first such measurements for these decay modes and among the world's most precise. $a_{CP}^{T-\mathrm{odd}}$ is also measured in sub-regions of phase space corresponding to the intermediate processes $D^+ \to \phi \rho^+$, $\Bar{K}^{*0}K^{*+}$ and $\Bar{K}^{*0}\rho^+$ as well as $D_s^+ \to K^{*+} \rho^0$, $K^{*0}\rho^+$, $\phi\rho^+$ and $\Bar{K}^{*0}K^{*+}$. All values are found to be consistent with zero. 

\begin{figure} [hbt!]
\centering
\includegraphics[width=0.65\textwidth]{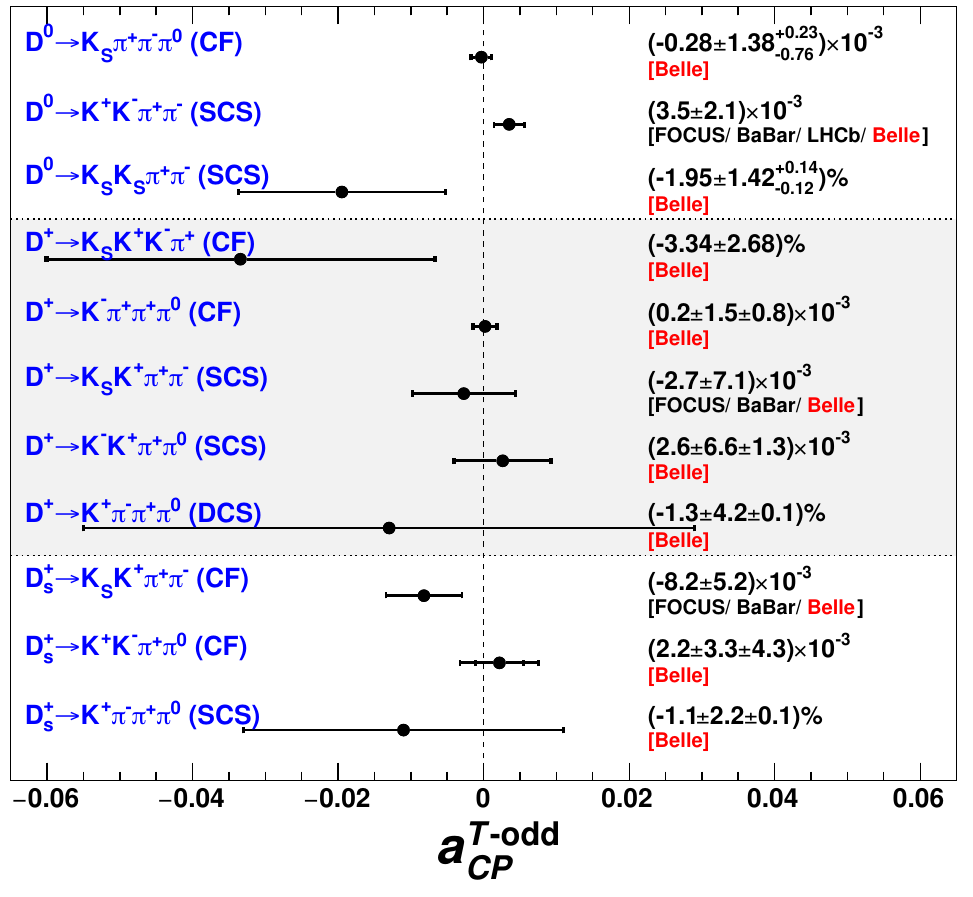}
 \caption{Overview of recent measurement for $a_{CP}^{T-\mathrm{odd}}$ including the ones presented here. \cite{llki}}
\label{fig:overview}
\end{figure}

A comprehensive overview of all the results including previous measurements is shown in Figure \ref{fig:overview}.


\section*{Acknowledgments}

The author would like to thank the Belle II Collaboration for the opportunity to present, and the organizers of the 12th Workshop on the CKM Unitarity Triangle for the successful conference. This work is supported by the Austrian Science Fund under award number J4625-N.

\bibliographystyle{amsplain}

\begin{thebibliography}{99}
    
\bibitem{cft}
I.~Adachi \textit{et al.} [Belle-II],
Phys. Rev. D \textbf{107} (2023) no.11, 112010

\bibitem{ksks}
N.~Dash \textit{et al.} [Belle],
Phys. Rev. Lett. \textbf{119} (2017) no.17, 171801

\bibitem{neuro}
M.~Feindt and U.~Kerzel,
Nucl. Instrum. Meth. A \textbf{559} (2006), 190-194

\bibitem{aman}
A.~Sangal \textit{et al.} [Belle],
Phys. Rev. D \textbf{107} (2023) no.5, 052001

\bibitem{moon}
H.~K.~Moon \textit{et al.} [Belle],
Phys. Rev. D \textbf{108} (2023) no.11, L111102

\bibitem{llki}
L.~K.~Li \textit{et al.} [Belle],
[arXiv:2305.12806 [hep-ex]].


\end{thebibliography}

\end{document}